\documentclass{elsarticle}
\usepackage{blindtext}
\usepackage{abstract}
\usepackage{pifont}
\usepackage{natbib}
\usepackage{geometry}
\usepackage{fleqn}
\usepackage{enumerate}
\usepackage{commath}
\usepackage[british]{babel}
\usepackage{amsmath}
\usepackage{amssymb}
\usepackage[hidelinks,linkcolor=black,urlcolor=black,citecolor=black]{hyperref} 
\usepackage{braket}
\usepackage[T1]{fontenc} 
\usepackage[utf8]{inputenc} 
\usepackage{pdfpages}
\usepackage[version=4]{mhchem}
\usepackage{soul} 
\usepackage{graphicx}
\usepackage{wrapfig} 
\usepackage{subcaption} 
\usepackage{float} 
\usepackage{booktabs}
\usepackage{epstopdf} 
\usepackage{cleveref} 

\begin{document}
\begin{frontmatter}

\title{Analytical model for release calculations \\ in solid thin-foils ISOL targets}

\author[SCKaddress,Polimiaddress]{L. Egoriti}
\author[SCKaddress,UCLaddress]{S. Boeckx}
\author[SCKaddress]{L. Ghys}
\author[SCKaddress,UGentaddress]{D. Houngbo\corref{correspondingauthor}}
\ead{donald.houngbo@sckcen.be}
\author[SCKaddress]{L. Popescu}

\cortext[correspondingauthor]{Corresponding author}

\address[SCKaddress]{Belgian Nuclear Research Centre (SCK•CEN), Boeretang 200, B-2400 Mol, Belgium}
\address[Polimiaddress]{Politecnico di Milano, Department of Energy, CeSNEF-Nuclear Engineering Division, Via Ponzio, 34/3, 20133 Milano, Italy}
\address[UCLaddress]{ICTEAM Inst., Univ. Catholique de Louvain, Louvain-la-Neuve, Belgium}
\address[UGentaddress]{Department of Flow, Heat and Combustion Mechanics, Gent University (UGent), St.-Pietersnieuwstraat 41, B-9000 Gent, Belgium}


\end{frontmatter}

\begin{abstract}
	A detailed analytical model has been developed to simulate isotope-release curves from thin-foils ISOL targets. It involves the separate modelling of diffusion and effusion inside the target. The former has been modelled using both first and second Fick's law. The latter, effusion from the surface of the target material to the end of the ionizer, was simulated with the Monte Carlo code MolFlow+. The calculated delay-time distribution for this process was then fitted using a double-exponential function. The release curve obtained from the convolution of diffusion and effusion shows good agreement with experimental data from two different target geometries used at ISOLDE. Moreover, the experimental yields are well reproduced when combining the release fraction with calculated in-target production.
\end{abstract}

\section{Introduction}
\label{sec:Introduction}

Isotope Separation On Line (ISOL) is a powerful technique to produce radioactive ion beams (RIB) which are used for experimental studies in many fields, including nuclear physics, astrophysics and nuclear medicine \cite{EURISOL:report_0509}.
\begin{figure} [H]
	\centering
	 \includegraphics[width=0.6\linewidth]{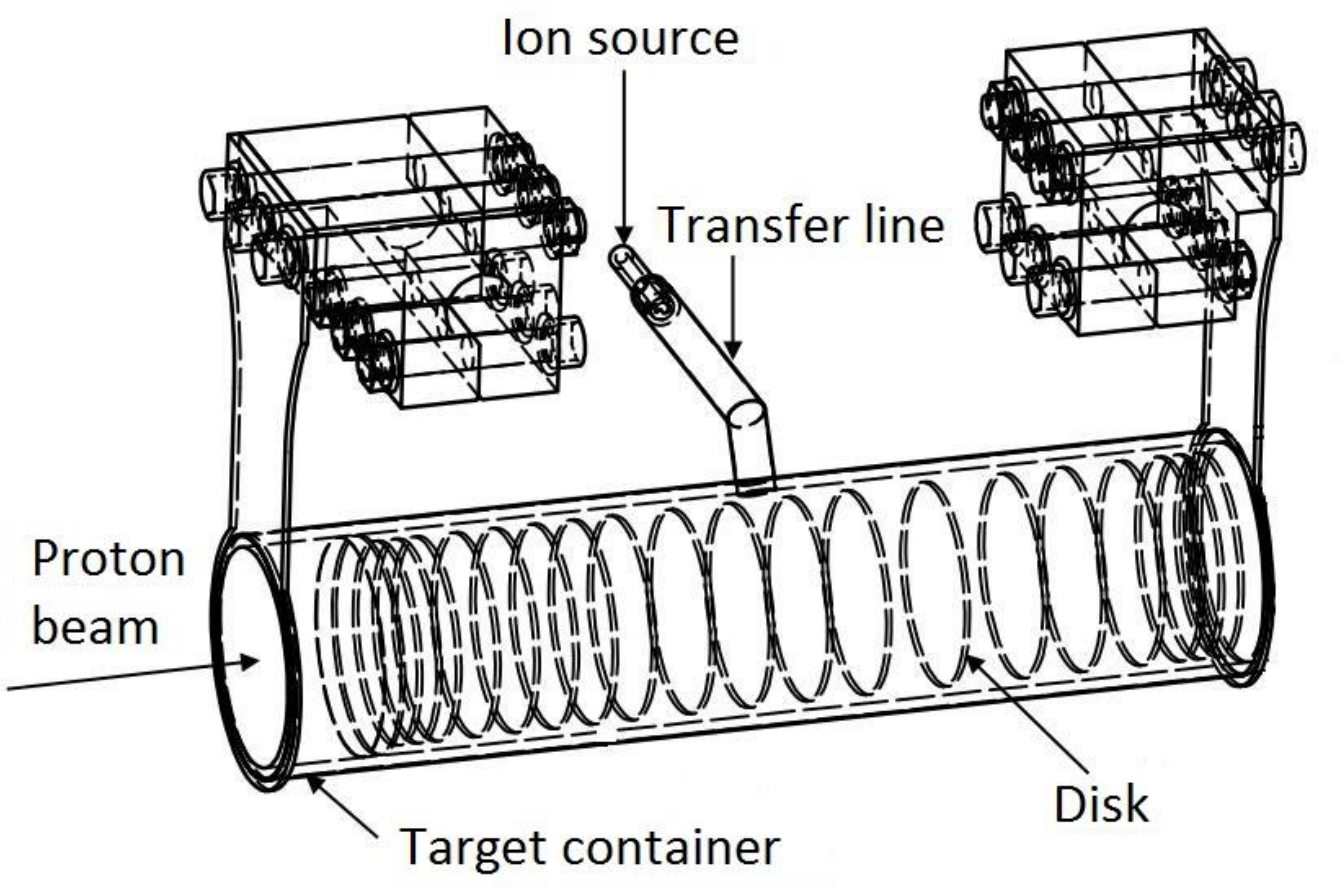}
	\caption{Schematic representation of a standard solid ISOL target. Picture extracted from \cite{Target:image}.}
	\label{fig:ISOL_Target}
\end{figure}
In the ISOL method isotopes are produced through the interaction of a high-energy light-particle beam (typically protons) with a high-Z target material, in which they are usually thermalized. These isotopes are subsequently released through diffusion in the material and effusion from the target-container volume towards an ion source where they are ionized and then extracted in an ion beam. Mass purification happens in a dedicated magnetic separator downstream the ion source.\\
Currently-used solid targets consist of a cylindrical container which is filled with pellets or thin foils, constituting the material in which radioactive nuclei are produced (see \cref{fig:ISOL_Target}).\\
Maximizing the RIB intensity at the experimental station is a major objective for any ISOL facility. This intensity $I$ is described by the following formula derived from \cite{VanDuppen:ISOL}:
\begin{equation}
	\label{eq:current}
	I = \int \phi(E,x) \sigma(E) \rho(x) N_{\rm A}/A  \,dE \,dx \ \epsilon_{\rm Diff} \ \epsilon_{\rm Eff} \ \epsilon_{\rm Ion} \ \epsilon_{\rm MS},
\end{equation}
where $\sigma(E)$ is the reaction cross section for the production of a specific isotope. This cross section depends on the target nuclide, with mass $A$, and the energy $E$ and type of the incoming particle. The target density is $\rho(x) $, while $N_{\rm A}$ denotes the Avogadro number. The other four terms in \cref{eq:current} represent the efficiencies related to diffusion out of the target material ($\epsilon_{\rm Diff}$), effusion through the target voids and transfer line ($\epsilon_{\rm Eff}$), ionization ($\epsilon_{\rm Ion}$) and mass separation ($\epsilon_{\rm MS}$).
An accurate description of the processes involved is crucial not only for predicting the outward isotope flux but also for target optimization. This paper reports on our approach where we model separately the in-target production, diffusion and effusion processes which are then combined to obtain an overall isotope-release curve in thin-foils targets.
The developed model is benchmarked with experimental data from two different solid tantalum-target geometries operated at ISOLDE and compared with previous calculations \cite{Bergmann:yields}.

\section{Methodology}
\label{sec:Methodology}

A first step towards target optimization for high RIB intensity is to accurately model the release of isotopes from the location where they are created to the ion source. In this section our release model for a solid thin-foils target is described. It includes diffusion inside the target material, effusion in the vacuum of both the target container and transfer line, and radioactive decay.

\subsection{Diffusion}
\label{sec:Diffusion}
The produced isotopes migrate from inside the solid material to the surface where they desorb. In the case of a thin-foils target, the thickness of each foil is much smaller than its other dimensions. Therefore, diffusion is only considered in the direction perpendicular to the surface of the foil. Fick's second law of diffusion \cite{Crank:diffusion}
\begin{equation}
	\frac{\partial C(\vec{x}, t)}{\partial{t}} = \nabla \cdot (D(\vec{x}) \nabla C(\vec{x},t))
\end{equation}
is thus simplified to
\begin{equation}
	\frac{\partial C(x, t)}{\partial{t}} = \frac{\partial}{\partial x}\left(D(x) \frac{\partial C(x,t)}{\partial x}\right)
\end{equation}
where $D$ is the diffusion coefficient of the migrating species in the material and $C$ represents their concentration profile.
The temperature dependence of the diffusion coefficient is usually described by an Arrhenius law \cite{Mehrer:diffusion}
\begin{equation}
	D = D_0 exp\left[-\frac{\Delta H}{RT}\right]
\end{equation}
where $T$ is the absolute temperature, $\Delta H$ is the activation enthalpy of diffusion and $R$ is the universal gas constant. This relation shows that the diffusion coefficient rises with increasing temperature. For this reason, an ISOL target has to be operated at a high temperature to enhance the diffusion of isotopes. In this study a uniform temperature distribution in the target material was assumed, thereby neglecting possible variations of the temperature profile due to beam-power deposition. Additionally, the initial concentration ($C_{\rm 0}$) of isotopes inside the target material is considered uniform.
These assumptions are crucial to obtain a linear problem, described by
\begin{equation}
	\frac{\partial C(x, t)}{\partial{t}} = D\ \frac{\partial^2  C(x,t)}{\partial x^2},
\end{equation}
which is solved analytically by the method of separation of variables, as shown in \cite{Crank:diffusion}. The solution is given by
\begin{equation}
	\label{eq:conc_profile}
	C(x,t) = \frac{4C_{\rm 0}}{\pi} \sum_{\rm n=0}^\infty \frac{1}{2n+1} exp\left[-\frac{(2n+1)^2 \pi^2 D}{l^2} t\right] \sin \left[\frac{(2n+1)\pi x}{l}\right],
\end{equation}
where $l$ is the foil thickness.
The flux of particles out of the solid target material $J$ is found by applying Fick's first law of diffusion \cite{Crank:diffusion}
\begin{equation}
	\label{eq:def_flux}
	J = - D \frac{\partial C}{\partial x} \bigg|_{x=l}.
\end{equation}
Inserting \eqref{eq:conc_profile} in \eqref{eq:def_flux} gives
\begin{equation}
	\label{eq:Diff_curve}
	J = \frac{4 C_0 D}{l} \sum_{\rm n=0}^\infty exp\left[-\frac{(2n+1)^2 \pi^2 D}{l^2} t\right].
\end{equation}

\subsection{Effusion}
\label{subsec:Effusion}
At the vacuum level of typical ISOL targets, molecular-flow conditions are satisfied \cite{MolecularFlow:Sn}. In this regime, an atom desorbed from a surface flies in a straight line until reaching another surface where it can be re-adsorbed.

Effusion, the migration of isotopes under these conditions, was simulated with the Monte Carlo code MolFlow+ \cite{MolFlow+:userguide}. The code tracks particles with mass $m$, starting from the target-material surface until they exit the ionizer. From MolFlow+ simulations, the distributions for the effusion delay-time and number of collisions can be derived. The Sticking time of particles on surfaces is not implemented in the code, which means that upon interaction the particle is either instantaneously re-desorbed or permanently trapped.\\
To correct for this effect, the mean sticking time $\tau$ of a particle to a surface is introduced in our model.

This parameter is related to the desorption rate $r_{\rm D}$ of a particle from a surface, given by (see e.g. \cite{Harald:SurfacesandInterfaces} )
\begin{equation}
\label{eq:Desorption_rate}
r_D = n \nu_0 exp\left[-\frac{E_{\rm act}}{k_{\rm B} T}\right],
\end{equation}
where $n$ is the number of adsorbate sites per area, $\nu_{\rm 0}$ is a frequency factor, $E_{\rm act}$ is the activation energy for desorption, $k_{\rm B}$ is the Boltzmann constant and $T$ is the absolute temperature.
The bonds between the particle and the surrounding material may be of chemical nature, in which case the activation energy is high, implying a low probability of being re-emitted. Alternatively, weaker van der Waals' interactions lead to a higher re-emission rate.\\
Since $r_{\rm D} \propto 1 / \tau$, one can write
\begin{equation}
\tau = \tau_0 exp\left[\frac{E_{\rm act}}{k_{\rm B}T}\right].
\label{eq:tau_T}
\end{equation}

Sticking time was included in the model by shifting the particle arrival time by
\begin{equation}
	\label{eq:DT_sticking}
	\delta t = t \left(\frac{\tau \overline{N_{\rm Coll}}}{\overline{t}} \right),
\end{equation}
where $\overline{t}$ and $\overline{N_{\rm Coll}}$ (both derived from MolFlow+ output) are respectively the average time which a particle has to travel before exiting the system and the average number of wall collisions.

Frequently, it is desirable to perform release calculations for various isotope masses or for different target temperatures. In this respect, note that both the mass $m$ and temperature $T$ are specified as input parameters in a MolFlow+ calculation. Under the condition that the effusion geometry is at a uniform temperature, the result of this calculation can be employed to derive a delay-time distribution for a different parameter set $m'$ and $T'$. This procedure, which avoids running multiple lengthy Monte Carlo simulations, is explained in the following paragraph.\\
In the MolFlow+ setting used in this work, a particle is re-emitted from a surface with the mean particle velocity $v$ according to the Maxwell-Boltzmann distribution \cite{MolFlow+:userguide}, given by
\begin{equation}
	\label{eq:Velocity_effusion}
	v = \sqrt{\frac{8 R T}{\pi m}}.
\end{equation}
Since the distance $d$ travelled by a particle to the end of the ionizer only depends on the target geometry one has
\begin{equation}
	\label{eq:Distance_travelled}
	d = v t = v' t' \rightarrow t' = \frac{v t}{v'},
\end{equation}
where $v$, $t$ and $v'$, $t'$ are respectively velocities and flight times corresponding to $m$, $T$ and $m'$, $T'$.\\
Substituting \cref{eq:Velocity_effusion} in \cref{eq:Distance_travelled} yields
\begin{equation}
	\label{eq:DTshift_mT}
	t' = \left(\sqrt{\frac{8R T}{\pi m}} \right) t \left(\sqrt{\frac{\pi m'}{8 R T'}}\right) = t \  \sqrt{\left(\frac{T}{m}\right) \left(\frac{m'}{T}\right)}.
\end{equation}

%
%
%
Applying the shift given by \cref{eq:DTshift_mT} to the MolFlow+ output results in the delay-time distribution for mass $m'$ and temperature $T'$.
The resulting discrete distribution, obtained after correcting for the sticking time (see \cref{eq:DT_sticking}), is then fitted with the function
\begin{equation}
	\label{eq:Eff_curve}
	E(t) = A (1-e^{-Bt})e^{-Ct}.
\end{equation}
A $\chi^2$-based method is employed to extract the parameters $A$, $B$ and $C$ from \cref{eq:Eff_curve}.

\subsection{Overall release}
The time evolution of the RIB intensity at the end of the ionizer $R(t)$ is modelled by the combination of diffusion, effusion and radioactive decay as
\begin{equation}
	\label{eq:Release_curve_definition}
	R(t) = e^{-\lambda t} \int_{\rm 0}^{t} J(t') E(t-t') dt'.
\end{equation}
Substituting \eqref{eq:Diff_curve} and \eqref{eq:Eff_curve} in \eqref{eq:Release_curve_definition} yields

\begin{equation}
	\label{eq:Release_curve}
	\begin{aligned}
		R(t)& = \frac{8C_0DA}{l} e^{- \lambda t} \left\{\sum_{\rm n=0}^{\infty}\left[\frac{exp\left(-\frac{(2n+1)^2 \pi^2 D}{l^2}t\right) - exp(-Ct)}{C- \frac{(2n+1)^2 \pi^2 D}{l^2}}\right] \right.\\
		& \quad \left.- \sum_{\rm n=0}^{\infty}\left[\frac{exp\left(-\frac{(2n+1)^2 \pi^2 D}{l^2}t\right) - exp(-(B+C)t)}{B+C- \frac{(2n+1)^2 \pi^2 D}{l^2}}\right]\right\}.
	\end{aligned}
\end{equation}

In addition, experimental yields can be compared with our model by introducing the release fraction $R_{\rm f}$ defined as
\begin{equation}
	\label{eq:Release_fraction}
	R_{\rm f} = \frac{\int_{\rm 0}^{\infty} R(t) dt}{\int_{\rm 0}^{\infty}\int_{\rm 0}^{t} J(t') E(t-t') dt' dt}.
\end{equation}
The RIB yield at the experimental station is computed by multiplying $R_{\rm f}$ with the isotope in-target production calculated in FLUKA \cite{FLUKA:code} \cite{FLUKA:usermanual} and the ionization efficiency $\epsilon_{\rm Ion}$.

\section{Results}
\label{sec:Results}

\begin{figure}[h]
	\begin{subfigure}{.5\textwidth}
		 \includegraphics[width=6cm,height=6cm]{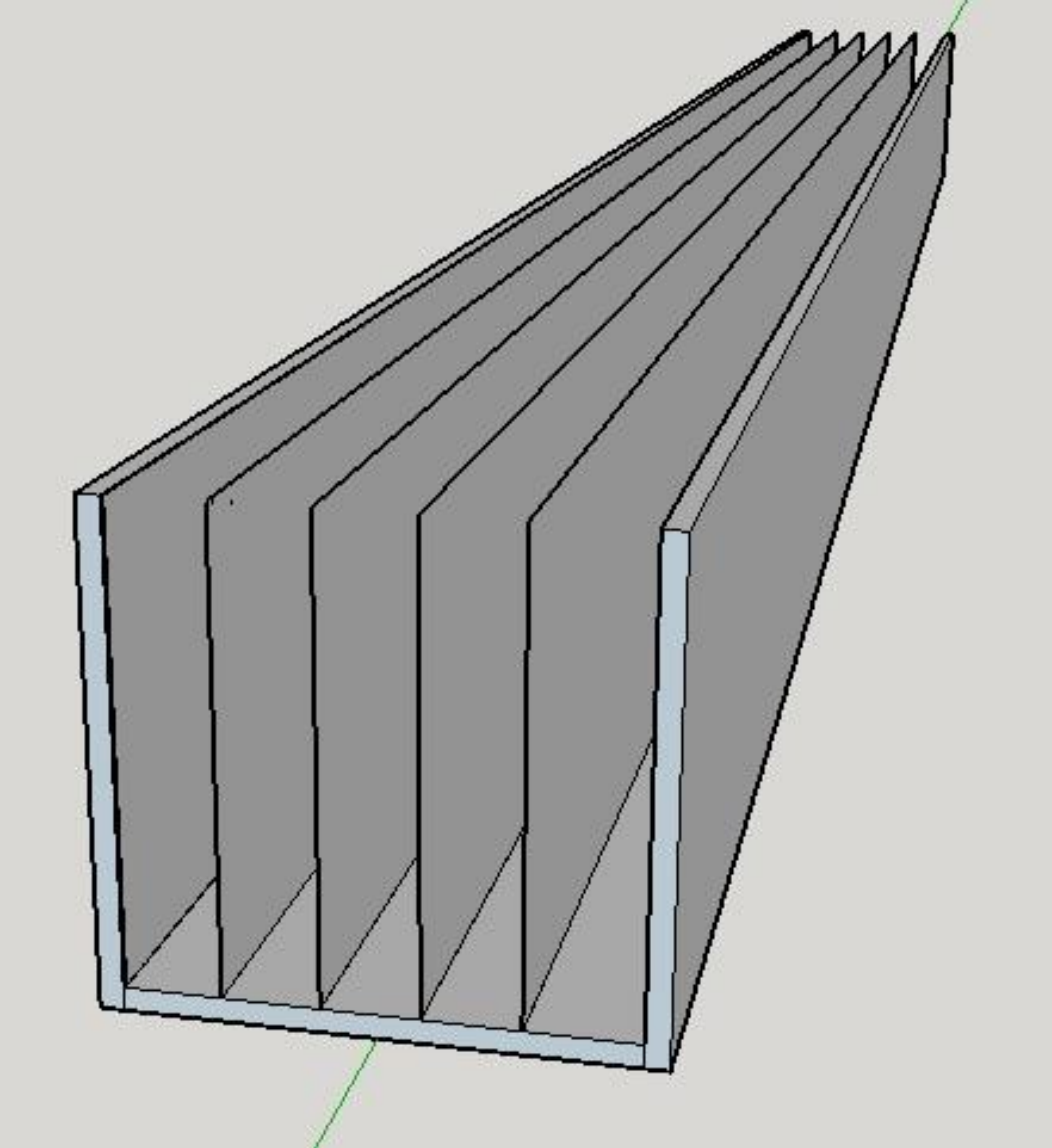}
		\caption{Ta129 target geometry.}
		\label{fig:Ta129_figure}
	\end{subfigure}%
	\begin{subfigure}{.5\textwidth}
		 \includegraphics[width=6cm,height=6cm]{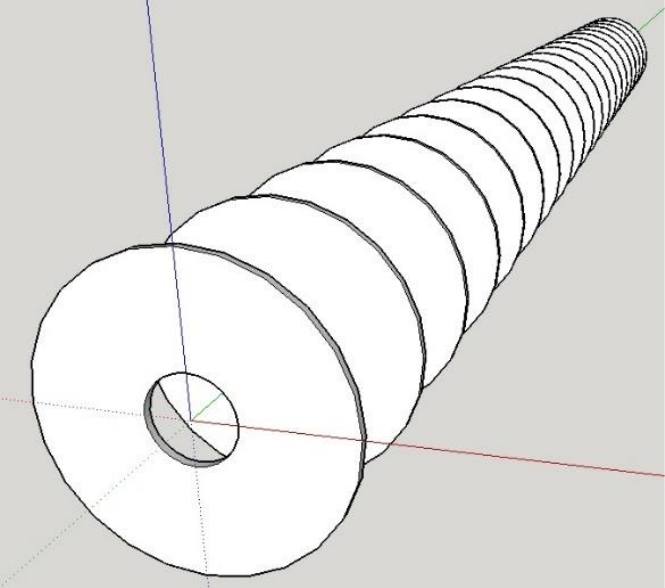}
		\caption{RIST-ISOLDE target geometry.}
		\label{fig:RIST-ISOLDE_figure}
	\end{subfigure}
	\caption{A representation of the different Ta-target geometries used for benchmarking our isotope-release model. The proton beam impinges parallel to the foils in geometry \ref{fig:Ta129_figure} and perpendicular in case \ref{fig:RIST-ISOLDE_figure}. The number of foils and annular discs is reduced for visualization purposes.}
	\label{fig:ISOLDE_targets_figure}
\end{figure}

This model is benchmarked with two different tantalum (Ta) targets operated at ISOLDE-CERN. In both cases, the target material is contained in a 20-cm long Ta cylinder with radius of 1 cm and operated at $T=2400 $ K \cite{Bennett:Li8}.
\begin{enumerate}
	\item The first geometry (\textit{Ta129} \cite{Bennett:Li8,Bennett:thinfoils,Mario:thesis} ) shown in a simplified way in \cref{fig:Ta129_figure} consists of 200 Ta foils (2-$\rm\mu$m thick, 15-cm long and 1-cm high) placed in the direction of the proton beam with 50-$\mu$m spacing between them.
	\item The second geometry (\textit{RIST-ISOLDE} \cite{Bennett:RIST_Li8,Mustapha:MCoptimization}) shown in \cref{fig:RIST-ISOLDE_figure} consists of 3600 Ta annular discs (25-$\rm\mu m$ thick, $\rm 9.5$\,mm external diameter and 2.5 mm internal diameter), placed perpendicular to the proton beam and spread over the entire length of the container.
\end{enumerate}

These two geometries have been selected because in both cases experimental release curves are available in literature \cite{Bennett:Li8,Bennett:RIST_Li8} for the same isotope ($\ce{^{8}Li}$).\\
At ISOLDE, protons arriving from the Proton Synchrotron Booster to the target station are grouped in pulses, separated by a multiple of 1.2\,s. Each proton pulse has a duration of maximum a few tens of\,$\mu$s, much smaller than typical diffusion or effusion times in the target ($\gtrsim 100$\,ms). Therefore, isotope formation was considered instantaneous in the current study. In addition, the release curve is calculated after a single proton pulse, consistent with literature data. In this work, isotopes are assumed to be created uniformly throughout the target material. This approximation is justified, in the discussed target geometries, since diffusion should only be considered perpendicular to the foils, and is thus independent of the particle concentration in the other directions. Additionally, no significant concentration variation is expected along the thickness of the few $\mu$m thin foils. In the effusion process, particles undergo on average several thousands of collisions, in each of which they are adsorbed and reemitted, thereby losing memory of their initial track. For this reason, also the effusion model is only marginally dependent on the initial distribution of isotopes in the target material.\\
As indicated in section \ref{sec:Diffusion}, the target is assumed to have a uniform temperature distribution. This approximation can be applied since local temperature variations due to beam-power deposition in the considered targets should be approximately one order of magnitude lower than the operating temperature of 2400\,K.\\
The mean sticking time $\tau$ of Li on a hot Ta surface is taken from literature and is equal to $2$ ns \cite{Mario:thesis,Eichler:sticking}, more than one order of magnitude below the mean flight time between two consecutive collisions for Li in these target geometries. Variations of $\tau$ due to the possible minor local temperature changes (see \cref{eq:tau_T}) can thus safely be neglected.
The diffusion coefficient $D$ for Li in Ta was not found in literature. As a consequence, $D$ was considered a free parameter in the comparison with the experimental release curve of $\ce{\rm ^{8}Li}$ in \textit{Ta129} (see \cref{fig:Ta129_matching}) and was determined by fitting the data. In the case of the RIST target, the theoretical normalized release curve was found to be much less sensitive to $D$ as compared to the Ta129-target geometry. With the current accuracy of the data, a reliable determination of $D$ through the same procedure was therefore prohibited. Ideally, the diffusion coefficient should be available in literature, in which case this model is applied without any fitting parameter.


\subsection{Thin-foils target: Ta129}
\begin{figure} [H]
	\centering
	 \includegraphics[width=0.6\linewidth]{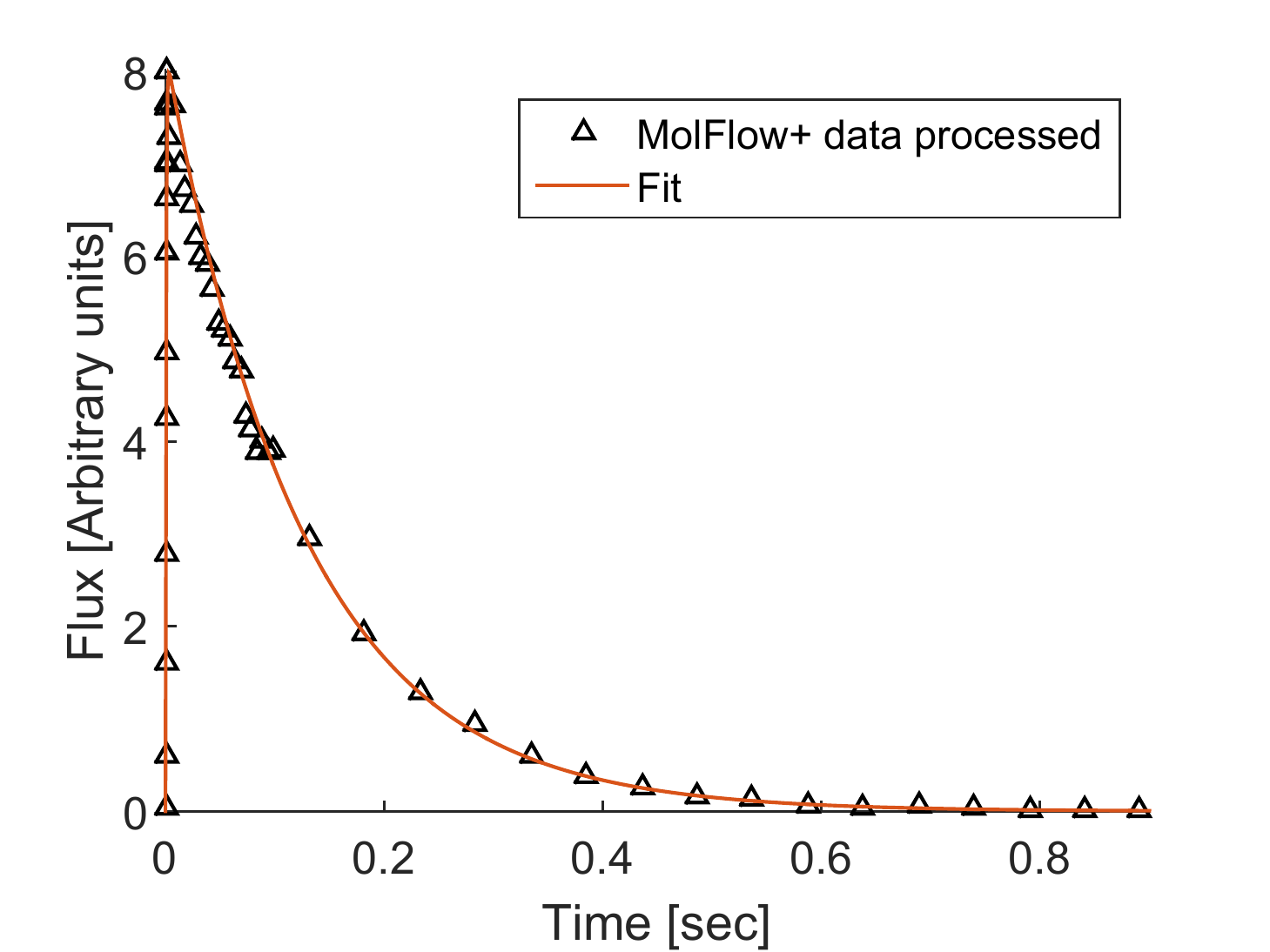}
	\caption{Effusion curve for \ce{\rm ^{8}Li} in Ta129 target (see \cref{fig:Ta129_figure}), calculated using MolFlow+. The solid line represents the best fit to the data using the function from equation \cref{eq:Eff_curve}, with $A = 8.2$, $B = 1.6 \cdot 10^3$, $C = 8.0$.}
	\label{fig:Eff_Ta129}
\end{figure}

The effusion curve for $\ce{\rm ^{8}Li}$ is obtained by fitting \cref{eq:Eff_curve} to MolFlow+ calculations (see \cref{subsec:Effusion}) with coefficients $A=8.2$, $B=1.6 \cdot 10^3$ and $C=8.0$, as shown in \cref{fig:Eff_Ta129}. These three parameters were then inserted in the expression of $R$, given by \cref{eq:Release_curve}.

\begin{figure} [H]
	\centering
	 \includegraphics[width=0.6\linewidth]{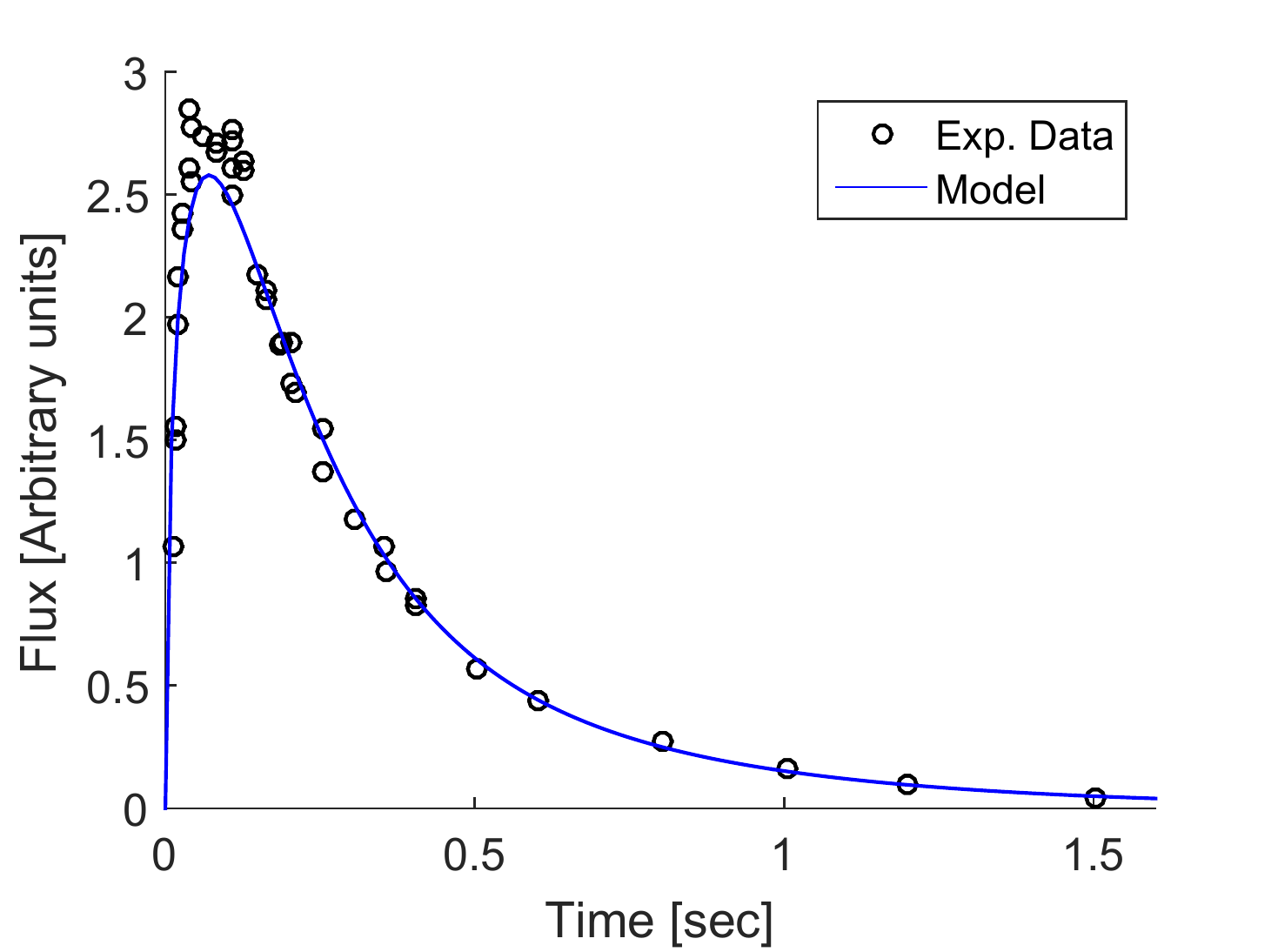}
	\caption{Normalized experimental release curve \cite{Bennett:Li8} of $\ce{^8Li}$ from Ta129 target, fitted with the calculated release curve yielding $D=5.4 \cdot 10^{-14}$\,$\mathrm{m^2 \ s^{-1}}$.}
	\label{fig:Ta129_matching}
\end{figure}

The experimental release curve taken from \cite{Bennett:Li8} was normalized in order to have its integral equal to unity, see \cref{fig:Ta129_matching}. The calculated release curve is normalized and fitted to the experimental data with a $\chi^2$ minimization procedure, assuming an equal relative weight for all fit points, which resulted in a value of $D=5.4(3) \cdot 10^{-14}$\,$\mathrm{m^2 \ s^{-1}}$.

\begin{figure}
\centering
	 \includegraphics[width=1.0\linewidth]{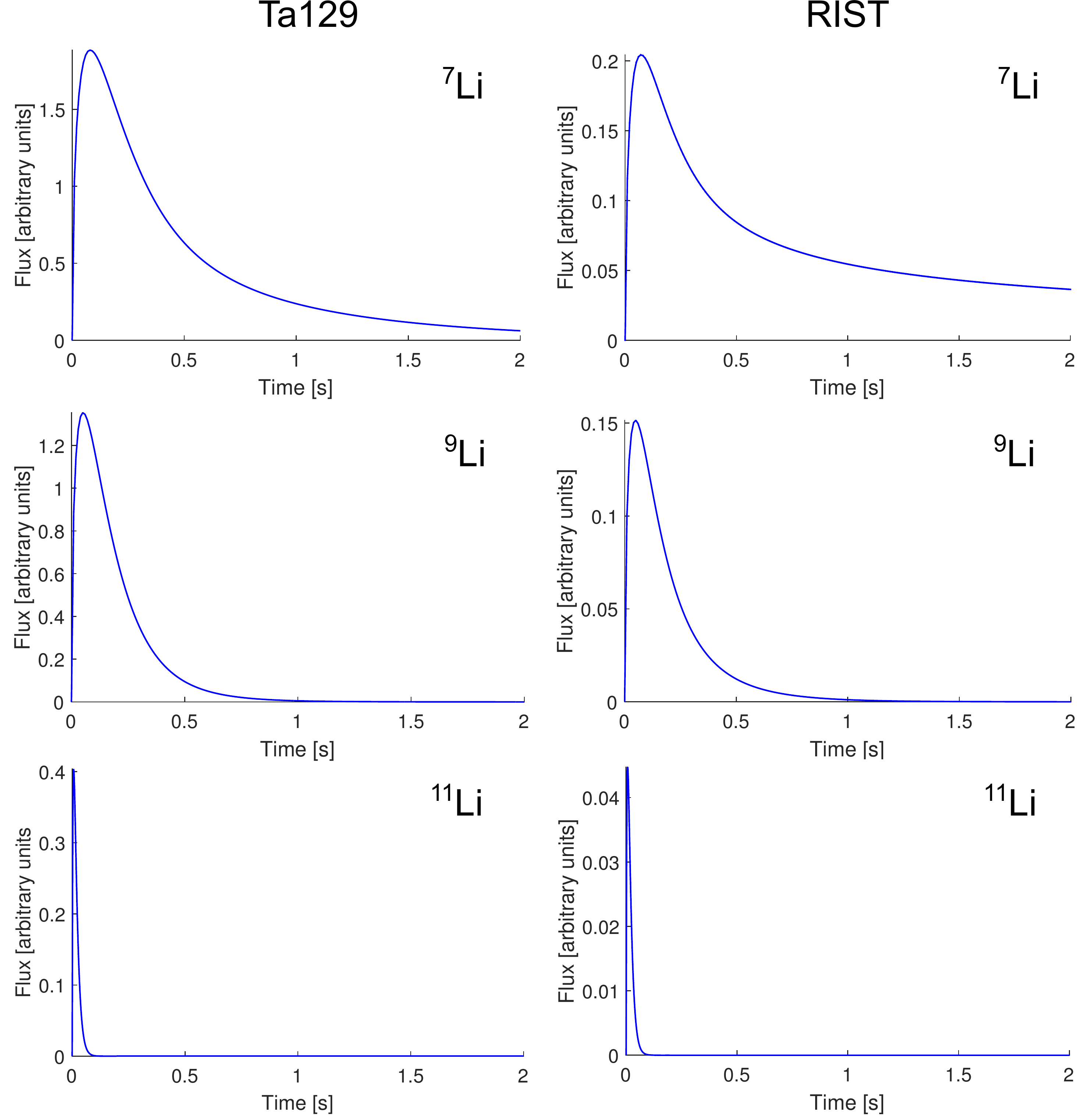}
	\caption{Predicted release curves for $^7$Li (stable), $^9$Li ($T_{1/2}=178$\,ms) and $^{11}$Li ($T_{1/2}=9$\,ms) for both the Ta129 (left) and the RIST (right) target geometries.}
	\label{fig:PredRelease}
\end{figure}

\begin{figure}
\centering
	 \includegraphics[width=0.6\linewidth]{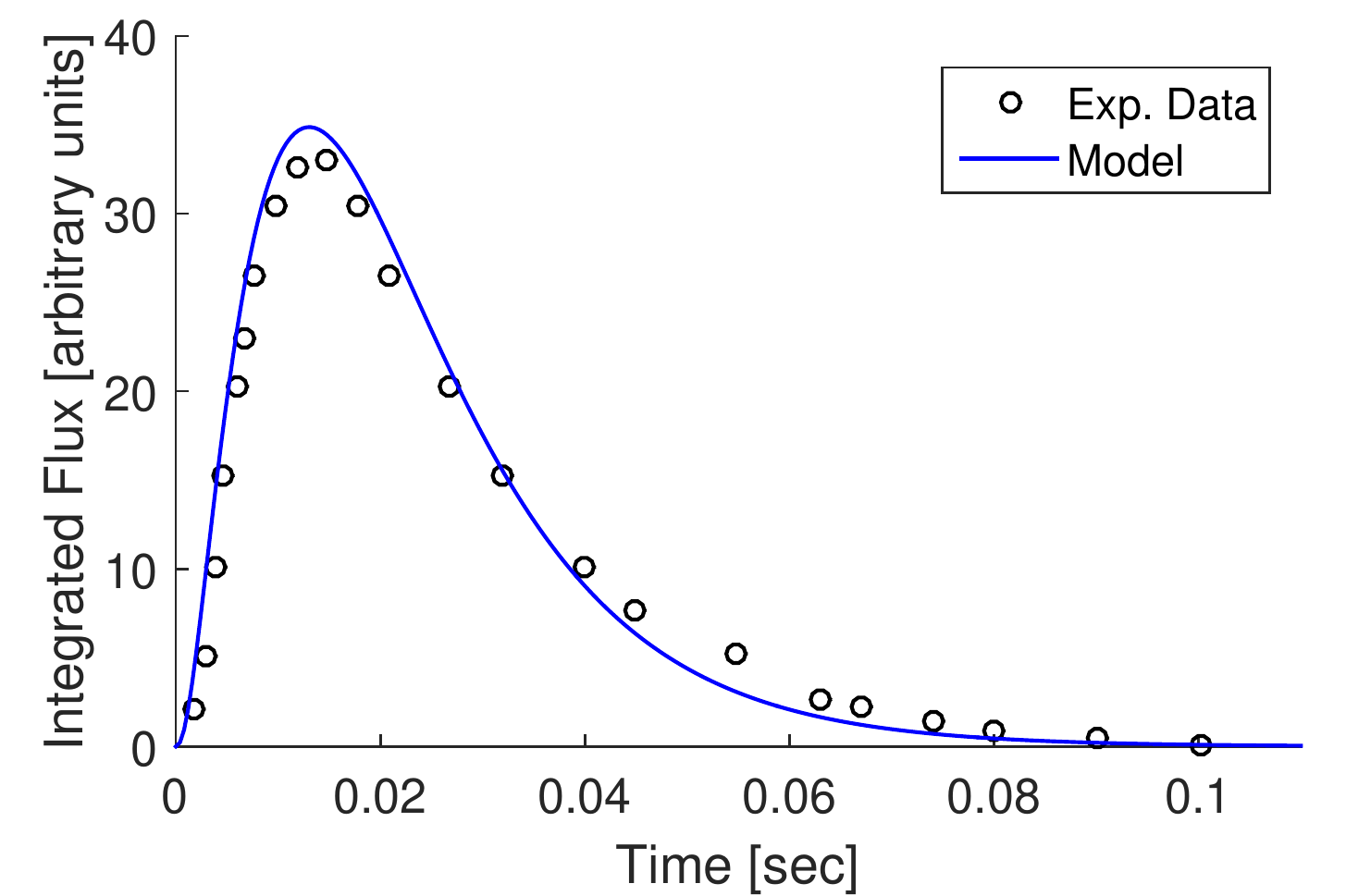}
	\caption{Comparison between the experimental \cite{Bennett:thinfoils} and theoretical integrated release curve from $^{11}$Li.}
	\label{fig:Li11_Comparison}
\end{figure}

Figure \ref{fig:Ta129_matching} shows reasonable agreement between the experimental points and the fitted curve, especially considering that only one free parameter $D$ was used. Nonetheless, a slight mismatch is observed at the cluster of points in the peak of the experimental distribution. This discrepancy can be attributed to an underestimation of the parameter $C$ in \cref{eq:Eff_curve}, describing the falling part of the MolFlow+ distribution.\\
Despite this small mismatch, \cref{tab:yields} shows a good agreement between experimental and calculated RIB yields of $\ce{Li}$ isotopes from \textit{Ta129}, using for the latter $\epsilon_{\rm ion} = 90$ \%, as reported in \cite{Bergmann:yields}. The diffusion coefficient $D=5.4(3) \cdot 10^{-14}$\,$\mathrm{m^2 \ s^{-1}}$, derived for $\rm ^8Li$, is applied in the calculations for all Li isotopes in \cref{tab:yields}. Figure \ref{fig:PredRelease} gives the predicted theoretical release curves, which were used to extract these value. In addition, \cref{fig:Li11_Comparison} shows a good agreement between the experimental \cite{Bennett:thinfoils} and calculated integrated release curve for $^{11}$Li. The conversion from the theoretical release curve shown in \cref{fig:PredRelease} to the integrated curve from \cref{fig:Li11_Comparison} is given by eq. (1) in reference \cite{Bennett:thinfoils}. The target geometry is taken from \cite{Bennett:thinfoils} and the diffusion coefficient was taken from the fit in \cref{fig:Eff_Ta129}. In this calculation, the half life of $^{11}$Li was taken as 9\,ms, equal to the value used in \cite{Bennett:thinfoils}.\\
Finally, \cref{tab:yields} shows that release fractions of our model agree reasonably well with previous results reported in \cite{Bergmann:yields}. Note that the approach followed in reference \cite{Bergmann:yields} is taken from \cite{Lettry:Release_model} and includes several phenomenological fitting parameters. It is therefore not applicable for target-design and optimization calculations.

\begin{table}[H]
	\caption{Comparison between experimental \cite{Bergmann:yields} and calculated (this study) yields. In-target production and corresponding statistical uncertainties, calculated with FLUKA, are also given. The release fractions from this model are compared with values \cite{Bergmann:yields} calculated from a phenomenological model.}
	\label{tab:yields}\centering
	\begin{tabular}{ c c c c c c c}
		\toprule
		Ion 			& $T_{\rm 1/2}$ & In-trg prod.  			 &	 $R_f$ 			& $R_f$						& 	Yield						 & Exp. Yield\\
		& 	[$ms$]		& $[1/\mu C]$ 				& (Our model)		& (\cite{Bergmann:yields})	& $[1/\mu C]$					 & $[1/\mu C]$\\
		\midrule
		$\ce{^{7}Li}$	& Stable		&  $3.14(1) \cdot 10^9 $	& $1$				& $1$						& $2.82 \cdot 10^9 $			&$2.0 \cdot 10^9$\\[1ex]
		$\ce{^{8}Li}$	& 839.9			&  $5.16(5) \cdot 10^8 $	& $0.68$			& $0.75$					& $3.17 \cdot 10^8 $			&$5.8 \cdot 10^8$\\[1ex]
		$\ce{^{9}Li}$	& 178.3			&  $9.6(2) \cdot 10^7 $	& $0.32$			& $0.45$					& $2.7 \cdot 10^7 $			&$1.7 \cdot 10^7$\\[1ex]
		$\ce{^{11}Li}$	& 8.75			&  $8.7(20) \cdot 10^5 $	&$0.01$				& $0.02$					& $7.6 \cdot 10^3 $			&$7.0 \cdot 10^3$ \\[1ex]
		\bottomrule
	\end{tabular}
\end{table}

\subsection{Annular-disc target: RIST-ISOLDE}
\begin{figure}[H]
	\centering
	 \includegraphics[width=0.6\linewidth]{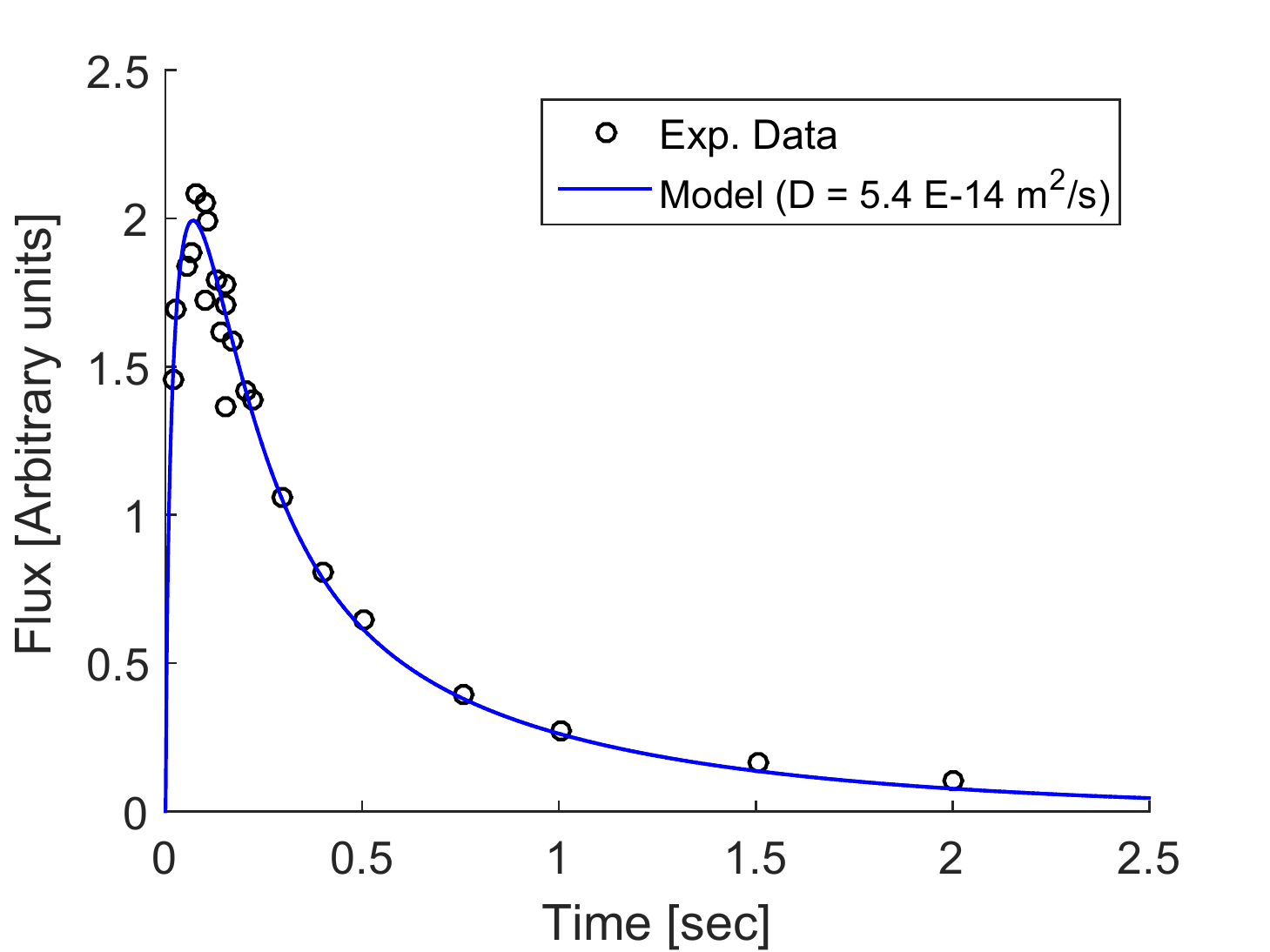}
	\caption{Release curve for RIST-ISOLDE target using the best-fit diffusion coefficient ($D=5.4 \cdot 10^{-14}$\,$\mathrm{m^2 \ s^{-1}}$) from the Ta129 geometry. The experimental data are taken from \cite{Bennett:RIST_Li8}.}
	\label{fig:Release_RIST}
\end{figure}

A similar methodology is applied to the RIST-ISOLDE target geometry. However,
as indicated before, the extraction of a reliable value of the diffusion coefficient was prohibited. Therefore, this parameter is taken equal to $D=5.4 \cdot 10^{-14}$\,$\mathrm{m^2 \ s^{-1}}$, as determined in the \textit{Ta129} case. As shown in \cref{fig:Release_RIST}, experimental data from \cite{Bennett:RIST_Li8} are well reproduced.
The effusion fit coefficients included in the calculation are $A = 12.3$, $B = 1.11 \cdot 10^3$ and $C = 11.2$. The release fraction of $\ce{\rm ^{8}Li}$ was found to be $9.6 \, \%$, close to the $\sim 8 \, \%$ reported in \cite{Mustapha:MCoptimization}, where the release from the \textit{RIST-ISOLDE} target is modelled using Geant-4 simulations. The calculated in-target production is $\rm 6.8 \cdot 10^9\,atoms/\mu C$, which leads to a RIB yield at the experimental station of $\rm 5.9 \cdot 10^8\,atoms/\mu C$ for $^{8}$Li, considering $\epsilon_{\rm Ion} = 0.9$. Predicted release curves for other Li isotopes are given in Figure \ref{fig:PredRelease}.

%
%
%


\section{Conclusion}
\label{sec:Conclusion}
An overall model for describing the release of isotopes out of solid thin-foils ISOL targets has been developed. This model combines the simplicity of an analytical description with the ability to simulate effusion in complex geometries. In-target production is simulated with \textit{FLUKA} while the diffusion process is calculated analytically. The effusion process is treated by means of MolFlow+ simulations, and fitted with an analytical function. The overall release curve was found by combining diffusion, effusion and radioactive decay.\\
This method allows calculating RIB yields for a wide variety of nuclides and temperature conditions for a certain geometry, without the need to re-run lengthy Monte Carlo simulations. Since no phenomenological fitting parameter is required, the model is ideally suited for the calculation of isotope release out of both existing and future targets.

\section*{Acknowledgements}
This work was partially funded by the GENTLE project and by the Belgian Science Policy Office through the BriX network project P7/12 of the Interuniversity Attraction Poles (IAP) programme.

\addcontentsline{toc}{section}{\refname}

\bibliography{Bibliography_elsevier}

\begin{thebibliography}{10}
\expandafter\ifx\csname url\endcsname\relax
  \def\url#1{\texttt{#1}}\fi
\expandafter\ifx\csname urlprefix\endcsname\relax\def\urlprefix{URL }\fi
\expandafter\ifx\csname href\endcsname\relax
  \def\href#1#2{#2} \def\path#1{#1}\fi

\bibitem{EURISOL:report_0509}
Y.~Blumenfeld, G.~Fortuna, Final report of the {EURISOL} {Design} {Study},
  Tech. rep., GANIL,
  \url{https://hal.inria.fr/file/index/docid/462950/filename/Final_Report_EURISOL_Design_Study.pdf}
  (2009).

\bibitem{Target:image}
{Tshoo, K.}, {Jang, D.Y.}, {Woo, H.J.}, {Kang, B.H.}, {Kim, G.D.}, {Hwang, W.},
  {Kim, Y.K.}, \href{http://dx.doi.org/10.1051/epjconf/20146611016}{Design
  study of 10 kw direct fission target for risp project}, EPJ Web of
  Conferences 66 (2014) 11016.
\newblock \href {http://dx.doi.org/10.1051/epjconf/20146611016}
  {\path{doi:10.1051/epjconf/20146611016}}.
\newline\urlprefix\url{http://dx.doi.org/10.1051/epjconf/20146611016}

\bibitem{VanDuppen:ISOL}
P.~Van~Duppen, Isotope separation on line and post acceleration,
  \url{http://euroschoolonexoticbeams.be/site/files/nlp/LNP700_contrib2.pdf}
  (2006).

\bibitem{Bergmann:yields}
U.~Bergmann, L.~Axelsson, J.~Bennett, M.~Borge, R.~Catherall, P.~Drumm,
  V.~Fedoseyev, C.~Forssén, L.~Fraile, H.~Fynbo, U.~Georg, T.~Giles,
  S.~Grévy, P.~Hornshøj, B.~Jonson, O.~Jonsson, U.~Köster, J.~Lettry,
  K.~Markenroth, F.~Marqués, V.~Mishin, I.~Mukha, T.~Nilsson, G.~Nyman,
  A.~Oberstedt, H.~Ravn, K.~Riisager, G.~Schrieder, V.~Sebastian, H.~Simon,
  O.~Tengblad, F.~Wenander, K.~W. Rolander,
  \href{http://www.sciencedirect.com/science/article/pii/S0375947401016116}{Light
  exotic isotopes: recent beam developments and physics applications at
  {ISOLDE}}, Nuclear Physics A 701~(1–4) (2002) 363 -- 368, 5th International
  Conference on Radioactive Nuclear Beams.
\newblock \href
  {http://dx.doi.org/http://dx.doi.org/10.1016/S0375-9474(01)01611-6}
  {\path{doi:http://dx.doi.org/10.1016/S0375-9474(01)01611-6}}.
\newline\urlprefix\url{http://www.sciencedirect.com/science/article/pii/S0375947401016116}

\bibitem{Crank:diffusion}
J.~Crank, The mathematics of diffusion, Oxford University Press, 1975.

\bibitem{Mehrer:diffusion}
H.~Mehrer, Diffusion in Solids, Springer, 2007.

\bibitem{MolecularFlow:Sn}
E.~G. Ronald, A study of the release properties of {Sn} and {SnS} from an
  isol-type target/ion source system, Master's thesis, University of Tennessee,
  Knoxville,
  \url{http://trace.tennessee.edu/cgi/viewcontent.cgi?article=1922&context=utk_gradthes}
  (2011).

\bibitem{MolFlow+:userguide}
R.~Kersevan, MolFlow+ user guide (2014).

\bibitem{Harald:SurfacesandInterfaces}
H.~Ibach, Physics of Surfaces and Interfaces, Springer, 2006.

\bibitem{FLUKA:code}
T.~Böhlen, F.~Cerutti, P.~Chin, A.~Fassò, A.~Ferrari, P.~Ortega, A.~Mairani,
  P.~Sala, S.~G., V.~Vlachoudis, The {FLUKA} code: Developments and challenges
  for high energy and medical applications, Nuclear Data Sheets 120 (2014)
  211--214.

\bibitem{FLUKA:usermanual}
A.~Ferrari, P.~R. Sala, A.~Fassò, J.~Ranft, FLUKA: a multi-particle transport
  code, $CERN-2005-10 (2005), INFN/TC_05/11, SLAC-R-773$ (2005).

\bibitem{Bennett:Li8}
J.~Bennett, U.~Bergmann, P.~Drumm, H.~Ravn,
  \href{http://www.sciencedirect.com/science/article/pii/S0168583X02019092}{The
  development of fast tantalum foil targets for short-lived isotopes}, Nuclear
  Instruments and Methods in Physics Research Section B: Beam Interactions with
  Materials and Atoms 204 (2003) 215 -- 219, 14th International Conference on
  Electromagnetic Isotope Separators and Techniques Related to their
  Applications.
\newblock \href
  {http://dx.doi.org/http://dx.doi.org/10.1016/S0168-583X(02)01909-2}
  {\path{doi:http://dx.doi.org/10.1016/S0168-583X(02)01909-2}}.
\newline\urlprefix\url{http://www.sciencedirect.com/science/article/pii/S0168583X02019092}

\bibitem{Bennett:thinfoils}
J.~R.~J. Bennett, U.~C. Bergmann, P.~V. Drumm, J.~Lettry, T.~Nilsson,
  R.~Catherall, O.~C. Jonsson, H.~L. Ravn, H.~Simon, Release studies of a thin
  foil tantalum target for the production of short-lived radioactive nuclei,
  in: Nuclear Physics A, 2002, pp. 327C--333C.

\bibitem{Mario:thesis}
M.~S. Leitner, A {Monte} {Carlo} code to optimize the production of radioactive
  ion beams by the {ISOL} technique., Ph.D. thesis, Technical University of
  Catalonia (UPC) (2005).

\bibitem{Bennett:RIST_Li8}
J.~R.~J. Bennett, et~al., {Progress of the RIST project}, Conf. Proc. C960610
  (1996) 1522--1524.

\bibitem{Mustapha:MCoptimization}
B.~Mustapha, J.~Nolen,
  \href{http://www.sciencedirect.com/science/article/pii/S0168583X02019262}{Optimization
  of {ISOL} targets based on monte-carlo simulations of ion release curves},
  Nuclear Instruments and Methods in Physics Research Section B: Beam
  Interactions with Materials and Atoms 204 (2003) 286 -- 292, 14th
  International Conference on Electromagnetic Isotope Separators and Techniques
  Related to their Applications.
\newblock \href
  {http://dx.doi.org/http://dx.doi.org/10.1016/S0168-583X(02)01926-2}
  {\path{doi:http://dx.doi.org/10.1016/S0168-583X(02)01926-2}}.
\newline\urlprefix\url{http://www.sciencedirect.com/science/article/pii/S0168583X02019262}

\bibitem{Eichler:sticking}
U.~Koester, Yields and spectroscopy of radioactive isotopes at {LOHENGRIN} and
  {ISOLDE}, Ph.D. thesis, TU Munchen (1999).

\bibitem{Lettry:Release_model}
J.~Lettry, et~al., Pulse shape of the isolde radioactive ion beams, Nuclear
  Instruments and Methods in Physics Research B 126 (1997) 130--134.

\end{thebibliography}

\clearpage

\end{document}